\documentclass[twocolumn,showpacs,preprintnumbers,amsmath,amssymb,natbib,noshowpacs]{revtex4-1}

\usepackage{graphicx}
\usepackage{dcolumn}
\usepackage{bm}
\usepackage{color}
\usepackage{hyperref}

\makeatletter
\def\l@subsection#1#2{}
\def\l@subsubsection#1#2{}
\makeatother

\begin{document}

\preprint{OU-HET-890}

\title{Topological Number of Edge States}

\author{Koji Hashimoto}
\email{koji@phys.sci.osaka-u.ac.jp}
 \affiliation{Department of Physics, Osaka University,
 Toyonaka, Osaka 560-0043, Japan.}
\author{Taro Kimura}%
 \email{taro.kimura@keio.jp}
\affiliation{%
Department of Physics, Keio University, Kanagawa
 223-8521, Japan.
}%
%


\begin{abstract}
We show that the edge states of the four-dimensional class A system can have topological charges, which are characterized by Abelian/non-Abelian monopoles.
 The edge topological charges are a new feature of relations among theories with different dimensions.
From this novel viewpoint, we provide a non-Abelian analogue of the TKNN number as an edge topological charge, which is defined by an $SU(2)$ 't Hooft--Polyakov BPS monopole through an equivalence to Nahm construction.
 Furthermore, putting a constant magnetic field yields an edge monopole in a non-commutative momentum space, where D-brane methods in string theory facilitate study of edge fermions.
\end{abstract}

\maketitle



\section{Introduction}
\label{sec1}

In the recent development in study of topological insulators \cite{hasan2010colloquium,qi2011topological},
the classification of the topological charges of the bulk states
by discrete symmetries and spatial dimensions
\cite{Schnyder:2008tya,Kitaev:2009mg} 
is widely used and provides a common ground for analysis of all continuum Hamiltonians.
Among the Hamiltonians, a particular important procedure is the dimensional reduction
\cite{Qi:2008ew,Ryu:2010zza}. One component of the momentum $p_i$ in the Hamiltonian 
is replaced by
a constant mass $m$, then the spatial dimensions reduce by one. The topological properties
may change along this procedure, but exhibit a universal reduction pattern. For example,
a class A topological insulator in four spatial dimensions, which is our main interest in this paper,
can be dimensionally reduced to a class AIII topological insulator in three dimensions.

The bulk-edge correspondence \cite{Jackiw:1975fn,hatsugai1993chern,Wen:2004ym} is the essential
viewpoint for
topological insulators both theoretically and experimentally. When the bulk wave functions of fermions possess a nontrivial topological number,
there appears a corresponding gapless edge states.
In view of the bulk-edge correspondence, it is natural to consider an alternative of the
dimensional reduction: introduction of a boundary. Generically, when a boundary is introduced
to a topological material with a nontrivial topological number in the gapped bulk,
there appear gapless edge states. At low energy, 
only the edge states can propagate and they exist only at the boundary, therefore
the spatial dimensions are reduced by one. This would serve as 
another way to realize a dimensional reduction. 

An interesting feature of this alternative dimensional reduction is that we have more freedom for possible reductions.
For example, the open boundary condition is typically applied to study the edge state.
However, the boundary condition satisfied by the fermions at the boundary is not unique: there appear a continuous family of boundary conditions. 
Furthermore, one can introduce more than a single boundary, say, parallel
two boundaries, each of which one can choose boundary conditions in principle.
Depending on
these details, the resultant edge states are different: they may have varieties of dispersions
and numbers of modes.

We would like to explore this alternative possibility for relating Hamiltonians in different dimensions.
Combining it with the bulk-edge correspondence,
we are naturally led to the idea of {\it 
topological charges carried by edge states}.
Normally the topological charges of the topological insulators are defined by the bulk states.
However, upon the dimensional reduction as giving the edge states, it would be natural to expect that some topological charges may show up also from the edge states.

In this paper, we analyze a class A topological insulator in four spatial dimensions, as one of 
the simplest examples. We discuss generic boundary condition {\it \`a la} Witten~\cite{Witten:2015aoa}, which is a different point of view from~\cite{Zhang:2012PRB,Enaldiev:2015JETP}, and choose a particular boundary condition which satisfies
the generic criteria for any consistent boundary condition of the system.

Our boundary condition is related to the mass term of the Hamiltonian.
With that choice, in this paper we find the followings:

\begin{itemize}
\item
For a single boundary, the Berry connection of the edge state provides a Dirac monopole.
The edge state is a gapless Weyl fermion in three dimensions, so, upon a normal dimensional
reduction to two dimensions (a massive two-dimensional fermion), the edge topological
charge is the same as that of the integer quantum Hall effect, that is, the TKNN number
\cite{Thouless:1982zz}.
\item For two boundaries which are parallel to each other, the Berry connection of the
two edge states is found to give a Berry curvature of a non-Abelian monopole. 
The monopole is the renowned 't Hooft--Polyakov monopole 
\cite{'tHooft:1974qc,Polyakov:1974ek}
in the BPS limit \cite{Bogomolny:1975de,Prasad:1975kr}. It would serve as a non-Abelian generalization of the TKNN number.
\end{itemize}

The emergence of the monopoles from the edge states is observed through our 
exact identification
of the Hamiltonian system with the Nahm construction of BPS monopoles \cite{Nahm:1979yw}.
The (non-)Abelian monopole charge of the edge states would be a new path for 
a characterization of topological insulators. The parallelism to the Nahm construction,
which is a method to exhaust all possible solution to the BPS monopole equation
for any gauge group and any monopole number, is expected to provide fertile applications
for more examples
and also a bridge to mathematical sciences.

Introduction of two boundaries in four spatial dimensions resembles
the domain-wall fermion formalism \cite{kaplan1992method,shamir1993chiral} which is
quite popular in lattice QCD.
See \cite{Kimura:2015ixh} for more explicit connection between this formalism and topological systems.
A difference from ours is just the boundary condition at
the boundaries, and we shall clarify this point. Other choice of the boundary conditions
would lead to more exploration of the topological structure of the edge states.

We also find that, once the whole system is put in a magnetic field, the (non-)Abelian monopoles
are replaced by BPS monopoles in a non-commutative space \cite{Hashimoto:1999zw,Gross:2000ss,Gross:2000ph,Gross:2000wc}. The non-commutative monopoles
have been studied in string theory as their natural realization is made by D-brane configurations.
We show that the effect of the magnetic field, interpreted by the slanted angle of the D-brane,
is reflected in the location of the fermions in the edge states.

The organization of this paper is as follows. In section \ref{sec2}, we shall give a review of
a two-dimensional (2D) class A topological insulator, and see that 
the edge states do not possess any topological number. Then in section \ref{sec3},
we consider a four dimensional class A topological insulator, and find that a single boundary
provides an edge state with a topological structure of a Dirac monopole, giving a TKNN
number. In section \ref{sec4}, we introduce two parallel boundaries to the system and
find that two associated edge states form a topological charge of an $SU(2)$  't Hooft--Polyakov monopole. We explain a difference to the domain-wall fermion formalism in
lattice QCD. In section \ref{sec5}, we introduce a magnetic field and show an equivalence
to monopoles in a non-commutative space, via a D-brane picture in string theory. 
Section \ref{sec6} is for our conclusion and discussions.


\section{Review of edge states of 2d class A topological insulator}
\label{sec2}

In this section, we show that the edge states of 
the class A topological insulator in two dimensions have a trivial Berry connection.
Since the bulk states are in two dimensions, the edge state is on a line and
has a wave function $\psi(p_1)$. The Berry connection of this edge state $\psi(p_1)$
is merely a single component
$A_1(p_1)$ which is always gauged away, thus it is obvious that there exist no
nontrivial Berry curvature for edge states of any 2D system.
Nevertheless here we review the 2D case since the example is
instructive in view of our main case of the four dimensions in the next section.

The Hamiltonian of the class A topological insulator in two dimensions is
\begin{align}
{\cal H} = p_1 \sigma_1 + p_2 \sigma_2 + m \sigma_3\, ,
\label{2dH}
\end{align}
where $\sigma_i$'s are the Pauli matrices.
The bulk dispersion relation is that of a relativistic particle with mass $m$,
\begin{align}
\epsilon = \pm \sqrt{(p_1)^2 + (p_2)^2 + m^2} \, . 
\label{bands}
\end{align}
The bulk system possesses
a nontrivial topological charge which is the renowned TKNN number \cite{Thouless:1982zz}. However
we are interested in possible topological charges of the edge states.

For construction of generic edge states of this system, we follow Witten's argument
\cite{Witten:2015aoa}.
Let us introduce a boundary at $x^2=0$, and consider a material in the region $x^2\geq 0$
only. Then we may generically impose a boundary condition
there,
\begin{align}
M\psi =-\psi \quad (x^2=0).
\label{Mb2}
\end{align}
Here $M$ is a generic Hermitian $2\times 2$ matrix~\footnote{
More generically $M$ needs not to be Hermitian. We will report a generic study elsewhere.
}.
Since at the boundary the Hamiltonian
needs to be self-conjugate, $\langle \psi_1|H\psi_2\rangle = \langle H \psi_1 | \psi_2\rangle$,
a partial integration over $x^2$ space provides a constraint:
\begin{align}
\{M,\sigma_2\} = 0 \, .
\end{align}
Such a matrix generically can be written as
\begin{align}
M = a_1 \sigma_1 + a_3 \sigma_3.
\end{align}
where $a_a$ and $a_3$ are real numbers. 
Any fermionic boundary condition kills
a half of the total components of the spinor, so $M$ needs to have a single $+1$ eigenvalue
and a single $-1$ eigenvalue, which means ${\rm tr}M=0$ and $\det M=-1$, resulting in
$a_1^2 + a_3^2 = 1$. So we may put $a_1=\cos\theta$ and $a_3 = \sin\theta$ for some $\theta
\in [0,2\pi)$.
Since our $p_1$ and $m$ appears $SO(2)$-symmetrically in the Hamiltonian (\ref{2dH}),
we are allowed to choose 
\begin{align}
M = \sigma_3,
\end{align}
then the boundary condition is
\begin{align}
\sigma_3 \psi + \psi = 0  \quad (x^2=0) \, .
\label{sigma3bc}
\end{align}

Let us derive an edge state. The Hamiltonian eigen equation is
\begin{align}
{\cal H} \psi = \epsilon \psi
\end{align}
which can be explicitly written with the two-component expression 
$\psi = (\xi,\eta)^{\rm T}$ as
\begin{align}
(m-\epsilon)\xi + (p_1 - \partial_2)\eta = 0 \, ,
\label{eq12}
\\
(p_1 + \partial_2)\xi - (m+\epsilon)\eta = 0 \, .
\label{eq22}
\end{align}
Here $\partial_2 \equiv d/dx^2$ is used instead of the momentum $p_2$
since we introduced the boundary $x^2=0$ and break the translational invariance.
Using the second equation to eliminate $\eta$ in the first equation, we arrive at
\begin{align} 
(m^2-\epsilon^2 + p_1^2 - \partial_2^2)\xi=0 \, .
\label{xieq2}
\end{align}
Since we are interested in the edge states which should exist between the bands (\ref{bands}),
we have a relation $\epsilon^2 < m^2 + p_1^2$. Then the generic solution of the 
differential equation (\ref{xieq2}) is
\begin{align}
\xi = & \xi_A \exp\left(x^2 \sqrt{m^2 + p_1^2- \epsilon^2}\right) 
\nonumber \\
& + \xi_B \exp\left(- x^2 \sqrt{m^2 + p_1^2- \epsilon^2}\right) \, .
\end{align}
The first term is non-normalizable in our region $x^2 \geq 0$, thus prohibited.
The second term solely cannot satisfy the boundary condition (\ref{sigma3bc}),
so, as a result, we need $\xi=0$ for all space. Plugging this into (\ref{eq12}) and 
(\ref{eq22}), we can solve them and obtain a dispersion for the state 
\begin{align}
\epsilon = -m
\end{align}
satisfied by
\begin{align}
\psi = {\cal N}(p_1)  \exp [p_1 x^2] 
\left(\begin{array}{c}
0 \\ 1
\end{array}
\right)
\, .
\label{edge2d}
\end{align}
This is the edge state.
The state exists only for $p_1<0$, otherwise the state is non-normalizable.
This is a kind of Fermi arc which appears in the edge dispersion.
The normalization ${\cal N}(p_1)$ can be fixed up to an arbitrary phase by
\begin{align}
1 = \int_0^\infty \! dx^2 \; \psi^\dagger \psi
\end{align}
which results in ${\cal N} = \sqrt{-2p_1}$.
The existence of the edge state is the consequence of the bulk-edge correspondence.

Now, let us consider a Berry connection of the edge state (\ref{edge2d}). It turns out that
the connection vanishes,
\begin{align}
A_1 \equiv i \int_0^\infty \! dx^2 \; \psi^\dagger \frac{d}{dp_1} \psi = 0 \, .
\label{A1-1}
\end{align}
So, the boundary edge state does not have
any topological structure~\footnote{We do not have any Wilson line, either.}.

It would be instructive to introduce two boundaries instead, at
$x^2 = \pm L$. We assume that at
both the boundaries the boundary conditions are the same and identical to (\ref{sigma3bc}).
Then, following the same steps, we reach a unique edge state
\begin{align}
\psi = \sqrt{\frac{p_1}{\sinh 2p_1 L}} \exp [p_1 x^2] 
\left(\begin{array}{c}
0 \\ 1
\end{array}
\right)
\, .
\label{edge2d2}
\end{align}
with the dispersion $\epsilon = -m$. 
One may wonder why we have only a single edge state while there are
two boundaries. In fact, we can find that
the boundary degrees of freedom is doubled, because the previous edge state (\ref{edge2d})
for a single boundary is valid only for $p_1<0$ while the present case (\ref{edge2d2})
is fine for any $p_1$.

The Berry connection of this edge state (\ref{edge2d2})
is calculated to vanish again,
\begin{align}
A_1 \equiv i \int_{-L}^L\! dx^2 \; \psi^\dagger \frac{d}{dp_1} \psi = 0 \, .
\label{A1-2}
\end{align}
So, there is no topological structure carried by the edge state, even if we introduce two
boundaries to the system.

We worked with the 2D Hamiltonian (\ref{2dH}), but it may be regarded as 
a Hamiltonian of a three dimensional (3D) massless fermion such as Weyl semimetals,
\begin{align}
{\cal H} = p_1 \sigma_1 + p_2 \sigma_2 + p_3 \sigma_3\, ,
\end{align}
related just by a dimensional reduction $p_3 = m$ \cite{Ryu:2010zza}. Then the
edge states 
(\ref{edge2d}) and (\ref{edge2d2}) propagate in the boundary two dimensions, 
with a linear dispersion relation $\epsilon = - p_3$.
We can calculate another component of
the Berry connection of the edge state, $A_3$, as well as the previous
$A_1$ (\ref{A1-1}) or (\ref{A1-2}). However, it again turns out that
they vanish,
\begin{align}
A_3 \equiv i \int\! dx^2 \; \psi^\dagger \frac{d}{dp_3} \psi = 0 \, .
\end{align}
Therefore, also in this case of three dimensions, the edge states do not carry any topological information.

From the next sections, we will find that the situation is different in higher dimensions.
In four-dimensional (4D) topological insulators, the edge states are found to carry nontrivial topological
numbers.

 \section{Dirac monopole from edge state in 4D class A topological insulator}
\label{sec3}

\subsection{4D class A system and a consistent boundary surface}

We start with a free class A system in four spatial dimensions, whose
Hamiltonian is provided by
\begin{align}
{\cal H} = \gamma_\mu p_\mu + \gamma_5 m
\label{4dtopH}
\end{align}
where $\mu=1,2,3,4$ is for the four spatial directions, and $m$ is the mass
of the fermion. Upon a dimensional reduction by one dimension, in other words, 
by replacing one of the momenta $p_3$ by another mass $m_3$, the system reduces
to a 3D class AIII topological insulator. This replacement is just
a renaming of the variable, so the following study will be applied also to
the 3D class AIII topological insulators.

We work with a familiar choice of the Clifford algebra
$\{\gamma_M, \gamma_N\} =2 \delta_{MN} {\bf 1}_4$ ($M,N=1,2,3,4,5$), 
\begin{align}
\gamma_\mu \equiv \left(
\begin{array}{cc}
0 &\bar{e}_\mu \\ e_\mu & 0
\end{array}
\right), \quad
\gamma_5 \equiv -\gamma_1 \gamma_2\gamma_3\gamma_4 
=
\left(
\begin{array}{cc}
{\bf 1}_2 &0 \\ 0& -{\bf 1}_2
\end{array}
\right),
\end{align}
with $e_\mu \equiv (i \sigma_i, {\bf 1}_2)$ and $\bar{e}_\mu \equiv (-i\sigma_i, {\bf 1}_2)$,
for $i=1,2,3$.
Using the Clifford algebra, it is easy to see that the Hamiltonian eigenvalue problem
in four dimensions
\begin{align}
{\cal H} \psi = \epsilon \psi
\label{4dH}
\end{align}
is solved by $\epsilon = \pm \sqrt{p_\mu^2 + m^2}$ which is a relativistic dispersion
relation of a particle in four spatial dimensions with the mass $m$. Upon the replacement $p_3$ by $m_3$, one can get a relativistic dispersion relation of a particle in three spatial dimensions with a mass $\sqrt{m^2 + m_3^2}$ \cite{Ryu:2010zza}.
The system enjoys the existence of a nontrivial second Chern class, and thus supports
a topological phase.

Let us introduce a boundary to this system. Suppose that at $x^4=0$ there exists a
boundary at which the system is terminated, and the material has a support only at
$x^4>0$. According to the bulk-edge correspondence, we expect a massless
edge state localized on the boundary surface $x^4=0$. In the subsequent sections, we shall
see how the edge state provides a topological charge given by a Dirac monopole.

First we seek for a consistent boundary condition put at $x^4=0$, by following a
general argument described for example in Ref.~\cite{Witten:2015aoa}.
A possible boundary
condition put at the boundary is expected to be of the form
\begin{align}
M \psi = - \psi
\label{bcM}
\end{align}
where $M$ is a Hermitian matrix~\footnote{Putting a Dirichlet boundary condition
$\psi=0$ at the boundary $x^4=0$, instead, results in no consistent on-shell
wave function. So we do not consider the Dirichlet boundary condition in this paper.
Our condition (\ref{bcM}) kills only a half of the fermion degrees freedom at the boundary,
which is natural.}.
We impose a self-conjugacy condition $\langle \psi_1 | H \psi_2\rangle
= \langle H\psi_1 | \psi_2\rangle$ for an arbitrary set of wave functions $\psi_1$ and $\psi_2$.
This Hermiticity condition is satisfied if the following property is met,
\begin{align}
\{M,\gamma_4\}=0
\label{Mgamma4}
\end{align}
since the partial integration over $dx^4$ involves $\gamma_4$ in the Hamiltonian. 
If we require
that the boundary condition (\ref{bcM}) is independent of the momentum of the 
fermion and demand the $SO(3)$ rotation invariance in the momentum space
$(p_1,p_2,p_3)$, we may choose a boundary condition~\footnote{
There exists another consistent boundary condition which does not break the $SO(3)$ 
symmetry: $M=\gamma_1\gamma_2\gamma_3$.}
\begin{align}
M = \gamma_5\, .
\label{bcmin}
\end{align}
Hence the boundary condition is
\begin{align}
\left(\gamma_5 + {\bf 1}_4\right) \psi \biggm|_{x^4=0} = 0 \, .
\label{bcx40}
\end{align}
In this paper, we consider this boundary condition, and will see the emergence of 
the monopole charge from the edge states.


\subsection{Spectrum and a unique edge state}

Now we solve the Hamiltonian eigen equation (\ref{4dH}) explicitly and find the 
edge state. Once the wave function $\psi$ is decomposed to $(\xi,\eta)^{\rm T}$
where $\xi$ and $\eta$ are two component spinors, the eigen equation is
\begin{align}
 (m-\epsilon)\xi + \left(\bar{e}_i p_i -i \frac{d}{dx^4}\right)\eta
 & = 0
\label{eq1}
\\
 \left(e_i p_i -i \frac{d}{dx^4}\right)\xi - (m+\epsilon)\eta
 & = 0
\label{eq2}
\end{align}
Note that $p_4$ is converted to a coordinate space $-id/dx^4$ so that we can treat the
boundary properly. The energy spectrum $\epsilon$ will be determined by 
the existence condition of the Hamiltonian eigen vectors.  
Multiplying $(m+\epsilon)$ on the first equation (\ref{eq1}), we can use the second equation (\ref{eq2})
to eliminate $\eta$, to obtain
\begin{align}
\left[p_i^2 - \left(\frac{d}{dx^4}\right)^2 - \epsilon^2 + m^2\right] \xi = 0 \, .
\end{align}
A generic solution reads
\begin{align}
\xi = \xi_A \exp[i\alpha x^4] + \xi_B \exp[-i\alpha x^4], \quad
\label{etaeq}
\end{align}
Here $\xi_A$ and $\xi_B$ are two-component spinors which are independent of
$x^4$, and $\alpha \equiv \sqrt{\epsilon^2-m^2-p_i^2}$. The solution is with a real $\alpha$ for $|\epsilon| \geq \sqrt{p_i^2 + m^2}$.

Let us impose the boundary condition (\ref{bcx40}). We obtain $\xi=0$ at $x^4=0$,
which amounts to the following constraint on $\xi$,
\begin{align}
\xi_A + \xi_B = 0.
\end{align}
Therefore a generic solution is
\begin{align}
\xi = 2 i \xi_A \sin(\alpha x^4)\, ,
\end{align}
with an arbitrary two-component spinor $\xi_A(p_i)$. The other component $\eta$ 
can be calculated from (\ref{eq2}) as
\begin{align}
\eta = \frac{2i}{m+\epsilon}\left(e_i p_i \sin(\alpha x^4)- i \alpha \cos(\alpha x^4)\right) \xi_A \, .
\end{align}
So, we find a family of solutions
parameterized by a two-component constant spinor $\xi_A$ and 
a real positive number $\alpha$ which is related to the energy as
\begin{align}
\epsilon = \pm \sqrt{p_i^2 + \alpha^2 + m^2} \, .
\label{epcont}
\end{align}
This $\alpha$ is a momentum along $x^4$. The dispersion is exactly the same
as that of the bulk state without the boundary. The positive energy is bounded from below
as $\epsilon \geq m$, and the system is gapped. 

On the other hand, for the other region of the energy, $|\epsilon| < \sqrt{p_i^2 + m^2}$,
we find a generic solution of (\ref{etaeq}) as
\begin{align}
\xi = \tilde{\xi}_A \exp[\tilde{\alpha} x^4] + \tilde{\xi}_B \exp[-\tilde{\alpha} x^4]\, .
\label{etaedgeb}
\end{align}
Here $\tilde{\alpha} \equiv \sqrt{-\epsilon^2+m^2+p_i^2}$ is 
a positive real constant. Since the material is defined in
a half-space $x^4>0$, the mode associated with $\tilde{\eta}_A$ is non-normalizable,
thus should not exist. So we need to consider only the mode
\begin{align}
\xi = \tilde{\xi} \exp\left[-\sqrt{-\epsilon^2 + m^2 + p_i^2} \; x^4\right] \, .
\end{align}
We impose the boundary condition (\ref{bcx40}), then this mode needs to
satisfy $\xi=0$ at $x^4=0$, which means 
\begin{align}
\xi = 0
\end{align}
for all space.
Using (\ref{eq2}), we find that this mode exists only at
\begin{align}
\epsilon = -m \, .
\label{em}
\end{align}
This flat band structure is similar to the Weyl semimetal surface state~\cite{Wan:2011PRB,Xu:2015Science}, but the current one is totally extend within the three-dimensional momentum space.
The remaining equation is (\ref{eq1}), 
\begin{align}
 \left(\bar{e}_i p_i -i \frac{d}{dx^4}\right)\eta
 & = 0.
\end{align}
Acting $(e_i p_i -i d/dx^4)$ on this equation leads to
\begin{align}
\left(p_i^2 - \left(\frac{d}{dx^4}\right)^2\right) \eta = 0.
\end{align}
A generic solution is $\exp (\pm px^4)$, whose sign is determined to be $-$
so that the mode is normalizable in the region $x^4\geq 0$. Thus we find
a generic solution
\begin{align}
\eta = \tilde{\eta} \exp[-px^4], \quad 
\left(p_i\sigma_i - p \right) \tilde{\eta}(p_i)=0,
\end{align}
%
%
%
with $p \equiv \sqrt{p_i^2}$. Using a unitary matrix $U(p_i)$ which diagonalizes
the matrix $p_i \sigma_i$ as
\begin{align}
p \sigma_3 = U^\dagger p_i \sigma_i U, 
\label{defU}
\end{align}
the spinor $\tilde{\eta}$ can be solved as
\begin{align}
\tilde{\eta} = U(p_i) \left(
\begin{array}{c}
1 \\ 0
\end{array}
\right)
\end{align}
Thus we are led to the following unique normalized solution,
\begin{align}
\eta = \eta_0(p_i) \equiv \sqrt{2p}\exp\!\left[- p x^4 \right] U(p_i) \left(
\begin{array}{c}
1 \\ 0
\end{array}
\right).
\label{edge1}
\end{align}
This mode is nothing but the edge state. The normalization is fixed by
\begin{align}
\int_0^\infty dx^4 \,\eta_0^\dagger \eta_0 = 1.
\label{norm1}
\end{align}

\begin{figure}[t]
\begin{center}
\includegraphics[height=5cm]{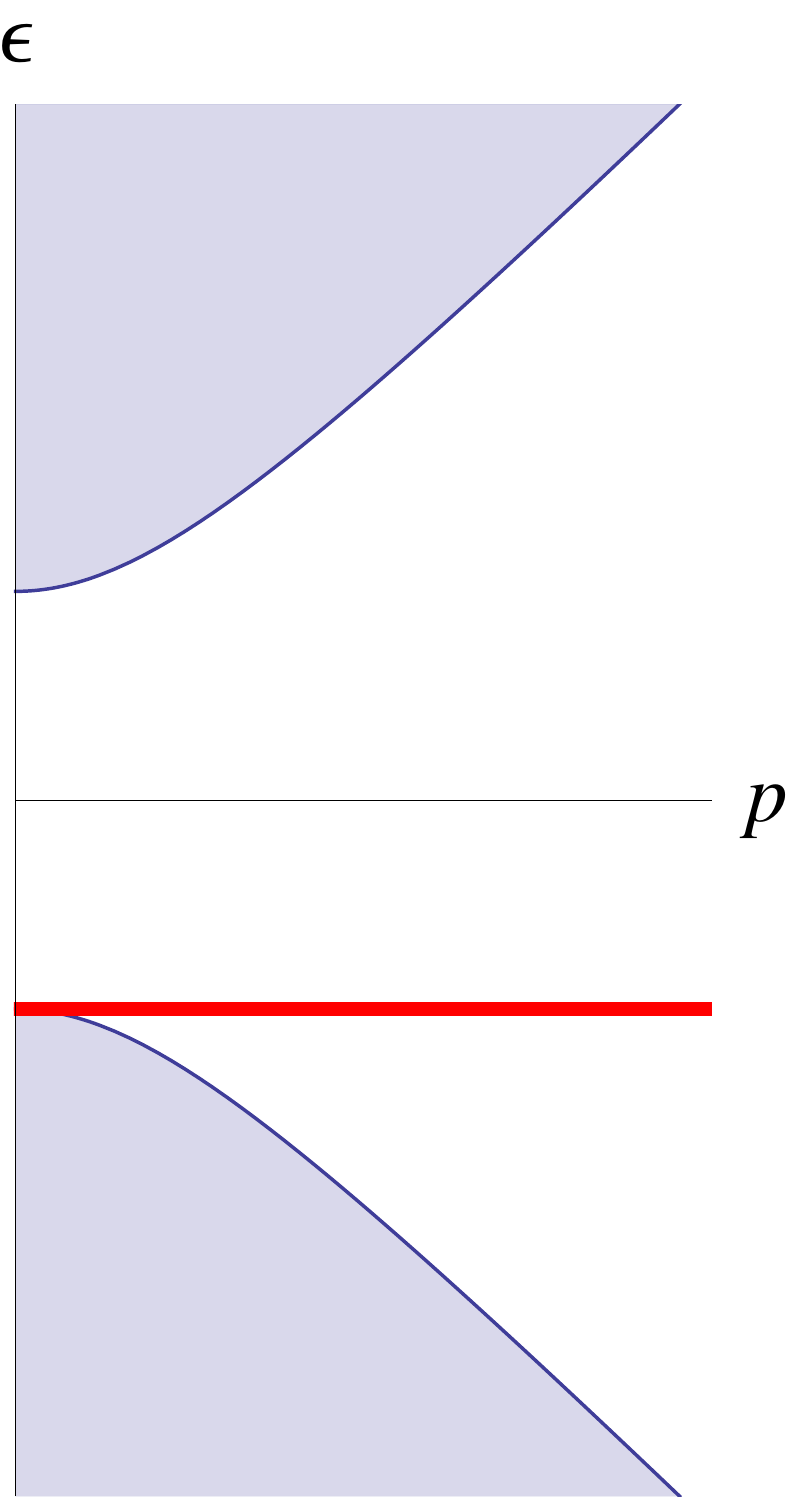}
\hspace{10mm}
\includegraphics[height=5cm]{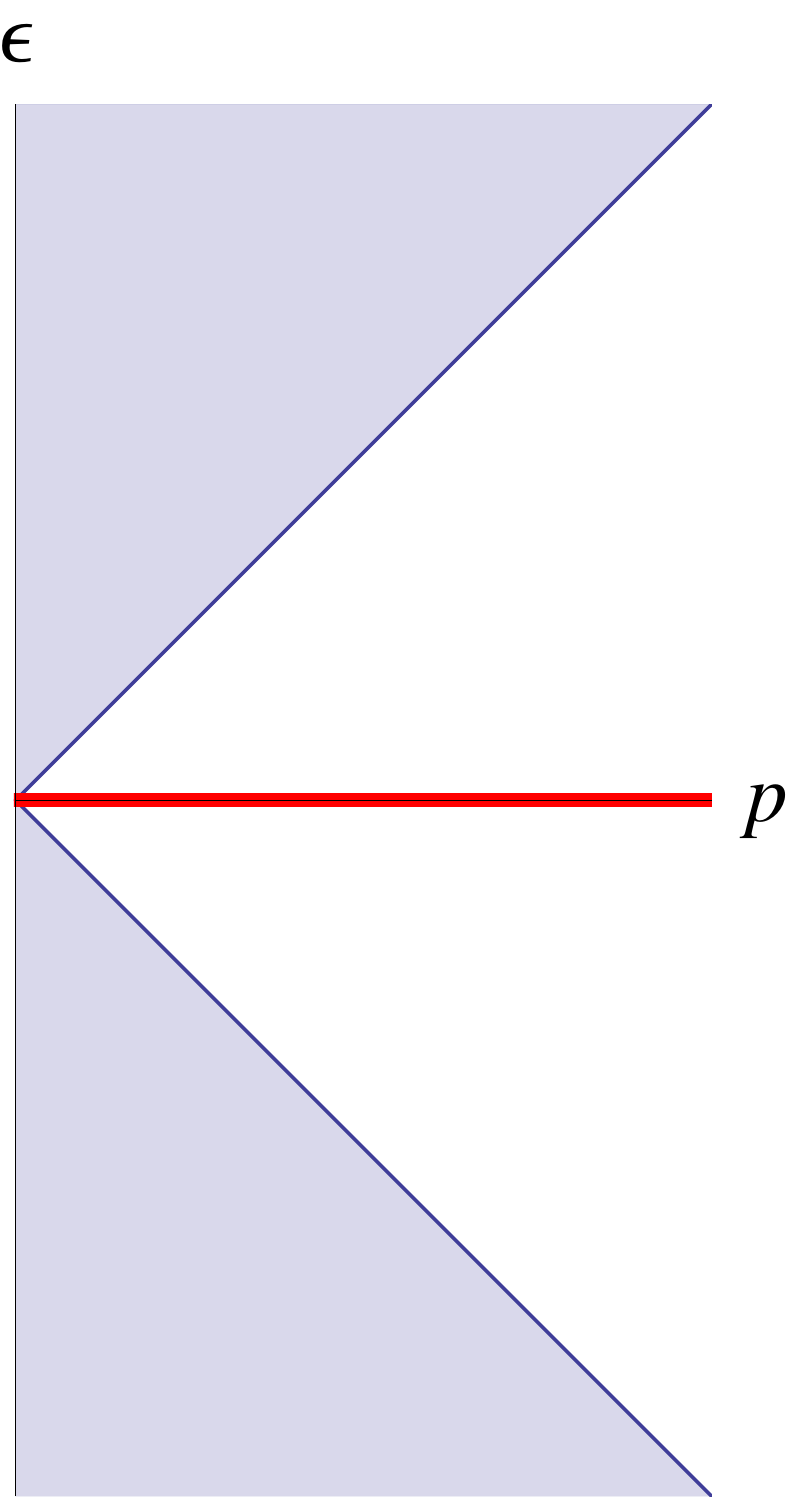}
\end{center}
 \caption{The obtained dispersions (\ref{dis1}) and (\ref{dis2}).
 Left panel is for $m\neq 0$,
 while 
 Right panel is for $m=0$.
 The red lines are for the edge state
 (\ref{dis2}) which is a flat band, and touches the tip of the bulk dispersion cone.
 }
\label{figdis} 
\end{figure}

So, in summary, we have obtained a bulk state and an edge state whose dispersion relations
are, respectively, given by (\ref{epcont}) and (\ref{em})
\begin{align}
&\epsilon = \pm \sqrt{p^2 + \alpha^2 + m^2} \quad \mbox{(bulk)} \, , 
\label{dis1}
\\
&\epsilon = -m  \quad \mbox{(edge)}\, .
\label{dis2}
\end{align}
The dispersion relations are illustrated in Fig.~\ref{figdis}.
Note that our surface state has a specific dispersion relation (\ref{em}), which 
does not depend on the momentum $p_i$, although the wave function itself
depends on $p_i$.
It is a generalization of a Fermi arc.
In fact, if we take a limit $m \rightarrow 0$, the energy of the 
edge state (\ref{em}) is $\epsilon = 0$ while the bulk dispersion (\ref{epcont})
becomes $\epsilon = \pm\sqrt{p^2 + \alpha^2}$. So the tip of the momentum
cone of the bulk dispersion coincides with the energy of the edge state.
In Weyl semimetals, generic Fermi arcs have a property that the arc is a flat dispersion
and ends at the Weyl points, and our case resembles that.
Topological aspects of flat bands are discussed, for example, in 
Refs.~\cite{Volovik:2010kj,Heikkila:2010qq}.

Next, we derive the topological charge of the edge state.

\subsection{Dirac monopole at the edge providing TKNN}

We shall see that a Berry connection associated with the boundary edge state $\eta_0(p_i)$ is identical to the Maxwell gauge field of a Dirac monopole. The Berry connection is
\begin{align}
A_i= \int_0^\infty dx^4 \; 
i \eta_0^\dagger \frac{d}{dp_i} \eta_0 \, .
\end{align}
Note that we need the integration over $x^4$ to define the Berry connection, 
since only under the integration
the self-adjointness of the operator $i d/dp_i$ follows, as
can be easily shown with the normalization condition (\ref{norm1})~\footnote{
If we didn't include the integration, the connection without the integration becomes complex. Nevertheless, it is worth being pointed out that the field strength is real
even without the integration.}.
One can think of this integration
as an integrated effect of the Berry connection, 
since the edge state is in any case almost localized at the boundary.
Explicitly, we find
\begin{align}
A_1 +i A_2 = \frac{i (p_1+ip_2)}{2p(p-p_3)}, \quad A_3=0 \, .
\end{align}
The field strength (the ``magnetic field'') calculated from this is
\begin{align}
B_i \equiv \frac12 \epsilon_{ijk}F_{jk} = \frac{p_i}{2p^3}
\label{1cc}
\end{align}
which provides a Dirac monopole configuration of a magnetic charge
\begin{align}
\frac{1}{2\pi}\int_{S^2} p^2 ds_i B_i
 = 1
\end{align}
where $ds_i$ is the volume element of the unit $S^2$.

It is instructive to evaluate the average location of the fermion perpendicular to the
boundary surface,
\begin{align}
\Phi(p_i)\equiv \int_0^\infty dx^4 \; \eta_0^\dagger \; x^4 \eta_0.
\end{align}
This measures the ``depth'' of the fermion for a given momentum.
In our case, an explicit calculation with the edge state (\ref{edge1}) leads to
\begin{align} 
\Phi = \frac{1}{2p}\,.
\end{align}
This means that the fermion with a momentum $p_i$ is located nearly
around $x^4 \sim 1/2p$. The smaller the momentum is, the deeper the
fermion lives from the boundary surface. 
The importance of this operator 
$\Phi$ is that it will be identified with a Higgs field for a symmetry breaking
of a U(2) connection for the case of two boundaries,
in the next section.

Let us point out a relation to the TKNN number \cite{Thouless:1982zz}, which 
will be important in the next section for defining a non-Abelian analogue
of the TKNN number. Suppose we tune the chemical potential around $\epsilon \sim -m$
and take a low energy limit. Then the bulk states disappear,
and we end up with only the edge state (\ref{edge1}).
Let us consider a standard dimensional reduction by replacing $p_3$ by a constant 
mass term $m_3$. Then the first Chern class (\ref{1cc}) is given by
\begin{align}
\nu = \frac{1}{2\pi}\int dp_1dp_2 \, F_{12} = \frac{1}{2}\mbox{sign} (m_3)
\label{TKNN}
\end{align}
This is the TKNN number of an integer quantum Hall system computed for the 2D Dirac Hamiltonian.

Here, to realize the 2D quantum Hall system, 
we look at only the edge states at $x^4=0$ of the 4D topological
insulator with the dimensional reduction $p_3=m_3$. 
This method provides us with a non-Abelian analogue of the TKNN number
in the next section.
The essence of the realization is 
that the state is provided by an edge state, rather than a bulk state.

Note that our energy dispersion is $\epsilon = -m$ is different from that of the
standard argument for the TKNN number, because we have started from four 
dimensions (\ref{4dtopH}) and considered the edge states, while the popular way
to get the TKNN number 
uses a bulk state, thus the Hamiltonians and spatial dimensions are different~\footnote{
In our previous work \cite{Hashimoto:2015dla}, we related the band spectrum with the shape 
of a D-brane in string theory. There we have used the energy band of the 3D
Weyl fermion, and the Hamiltonian was different. So, in the present case, there is no explicit and 
direct 
relation between the energy band of our edge state (which is a flat band) and the shape 
of a D-brane.}.

\section{'t Hooft--Polyakov monopole from edge states}
\label{sec4}

In this section, we shall show that the 4D topological insulator of class A
with two parallel boundaries
has a novel topological charge: a non-Abelian 
't Hooft--Polyakov monopole as its Berry connection of edge states.

\subsection{4D Topological insulator with two parallel boundaries}

We introduce two boundary surfaces which are parallel to each other.
The surfaces of the four dimensional topological insulator are three-dimensional, and defined by
$x^4 = \pm L$. The material exists between the two parallel surfaces, $-L \leq x^4 \leq L$.

As has been studied, a consistent boundary condition for the fermion is 
$(\gamma_5 \pm {\bf 1}_4)\psi=0$. We choose the same boundary 
conditions for both of the two boundary surfaces,
\begin{align} 
(\gamma_5+{\bf 1}_4)\psi\biggm|_{x^4=\pm L} = 0 \, .
\label{bdry2}
\end{align}
The choice is made 
just for getting the non-Abelian monopole for our purpose. We shall later discuss
other boundary conditions.

When the mass $m$ of the Hamiltonian ${\cal H} = \gamma_\mu p_\mu + \gamma_5 m$
is smaller than the scale coming from the width of the insulator $2L$, the insulator is a
``thin'' insulator.
Note that the thin limit is different from the dimensional reduction studied generally in Ref.~\cite{Ryu:2010zza}.

As in the previous section, we calculate the spectra of the fermion with the boundary condition (\ref{bdry2}). The fermion is decomposed again as $(\xi,\eta)^{\rm T}$, then
the boundary condition (\ref{bdry2}) is equivalent to say $\xi=0$ at $x^4=\pm L$.
The generic solution at the continuum (\ref{etaeq}) now needs to obey
the boundary condition 
\begin{align}
\xi_A\exp[\pm i\alpha L] + \xi_B\exp[\mp i \alpha L] = 0 \, .
\label{abexp}
\end{align}
This equation has a nontrivial solution only 
for special values of $\alpha$,
\begin{align} 
\alpha  = \frac{\pi}{2L} n \, ,  \quad n \in \mathbb{Z}
\end{align}
For these values of $\alpha$, arbitrary constant spinors
$\xi_A$ and $\xi_B$ satisfying
\begin{align}
\xi_A + \xi_B (-1)^n=0
\end{align}
are a solution. The other component $\eta$ can be constructed by
solving (\ref{eq2}). So we arrive at a mode with a dispersion relation
%
%
%
%
\begin{align} 
\epsilon = \pm \sqrt{p_i^2 + \frac{\pi^2n^2}{4 L^2}+m^2} \, .
\end{align}
We find that the continuous states (\ref{epcont}) is now discretized to
Kaluza--Klein states labeled by the integer $n$.

Next, let us look at the case (\ref{etaedgeb}). Imposing the boundary condition, we obtain
\begin{align} 
\tilde{\xi}_A \exp[\pm \tilde{\alpha} L] + 
\tilde{\xi}_B \exp[\mp\tilde{\alpha} L] = 0 
\end{align}
with, again, $\tilde{\alpha} \equiv \sqrt{-\epsilon^2+m^2+p_i^2}\geq 0$. 
The unique solution of this equation for a generic $p_i$ is $\xi_A=\xi_B=0$. Therefore, we conclude 
$\xi=0$ for all space.
Using (\ref{eq2}), we find the flat dispersion relation
\begin{align}
\epsilon = -m\, .
\label{e-m}
\end{align}
Using (\ref{eq1}) multiplied by $(e_i p_i -id/dx^4)$ 
with $\xi=0$, we find a generic solution
\begin{align}
\eta = \tilde{\eta}_A \exp[px^4]
+\tilde{\eta}_B \exp[-px^4].
\end{align}
%
%
Again using (\ref{eq1}) itself, we find that the mode needs to satisfy
\begin{align} 
(\bar{e}_i p_i - i p)
\tilde{\eta}_A \exp[px^4]+
(\bar{e}_i p_i + i p)
\tilde{\eta}_B \exp[-px^4]= 0 \, . 
\end{align}
For this to be satisfied at arbitrary $x^4$, 
\begin{align} 
(\bar{e}_i p_i - i p)
\tilde{\eta}_A =
(\bar{e}_i p_i + i p)
\tilde{\eta}_B= 0 \, . 
\end{align}
Then we obtain a generic solution with the dispersion (\ref{e-m}),
\begin{align} 
\eta = & c^+(p_i)\eta^+ 
+c^-(p_i) \eta^{-}\, ,
\\
\xi= & 0\, ,
\end{align}
where $c^\pm(p_i)$ are arbitrary complex functions of $p_i$, 
and 
%
\begin{align} 
&\eta^{+}\equiv
\sqrt{\frac{p}{\sinh 2pL}}
\exp[px^4] \; U(p_i)\left(
\begin{array}{c}
0 \\ 1 \end{array}\right) \, , 
\label{edgenon2}
\\
&\eta^{-} 
\equiv \sqrt{\frac{p}{\sinh 2pL}}
 \exp[-px^4] \; U(p_i)\left(
\begin{array}{c}
1 \\ 0 \end{array}\right) \, .
\label{edgenon}
\end{align}
These modes satisfy the eigen equation and the ortho-normalization condition
\begin{align}
& \left(\frac{\partial}{\partial x^4}+ p_i \sigma_i\right) \eta^\pm =0\, , 
\label{pipeta}\\
& \int_{-L}^L dx^4 \; (\eta^a)^\dagger \eta^{b} = \delta_{ab} \quad (a,b=\pm) \, .
\label{ortho}
\end{align}
$U(p_i)$
is the unitary matrix defined in (\ref{defU}), and explicitly given by
\begin{align}
U(p_i)=\frac{1}{\sqrt{2p(p-p_3)}}
\left(
\begin{array}{cc}
p_1-ip_2 & p_3-p \\
p-p_3 & p_1 + i p_2
\end{array}
\right)\, .
\end{align}
The states (\ref{edgenon2}) and (\ref{edgenon}) are edge states.
They are localized mostly at different boundary surfaces:
$\eta^\pm$ is localized at $x^4=\pm L$. 

We have two edge states sharing exactly the same dispersion
relation, as we have introduced two boundaries of the same boundary condition.
The number of degrees of freedom is doubled, compared to the case
of the single boundary: the situation is similar to that of the 2D case
in the previous section.

In summary, we have obtained the full spectra
\begin{align}
\epsilon &= \pm \sqrt{p^2 + \frac{\pi^2n^2}{4 L^2} + m^2} \quad \mbox{(bulk)} \, ,
\\
\epsilon &= -m  \quad \mbox{(two edge states)}\, .
\end{align}
We are interested in the edge states.
Tuning the chemical potential around $\epsilon \sim -m$ and taking a low energy limit,
the edge states play a dominant role. Two edge states satisfy (\ref{pipeta}) which 
can be thought of as a doubled pair of the edge state considered in the previous section.
Using these two edge states, we construct a non-Abelian charge and analyze an analogue of the TKNN number, in the next subsection.

\subsection{'t Hooft--Polyakov monopole from edge states}
 
Using the edge states (\ref{edgenon2}) and (\ref{edgenon}), we define a $2\times 2$ matrix-valued 
non-Abelian Berry connection~\footnote{See for example~\cite{hatsugai2004explicit,hatsugai2005characterization,hatsugai2006quantized}
for a non-Abelian Berry connection.
}  
\begin{align} 
A_i^{ab} \equiv i \int_{-L}^L dx^4 \; (\eta^a)^\dagger 
\frac{d}{dp_i} \eta^b \, .
\label{Ai}
\end{align}
Here $a,b=\pm$ labels the two edge states. 
In addition, we define a matrix-valued scalar operator $\Phi$ 
\begin{align} 
\Phi_i^{ab} \equiv  \int_{-L}^L dx^4 \; (\eta^a)^\dagger 
x^4 \eta^b \, .
\label{Phi}
\end{align}
This $\Phi$ measures the location of the fermion in the $x^4$ direction for a given
momentum $p_i$, for each $\eta^+$ and $\eta^-$ boundary edge state. Note that
the ``location'' has off-diagonal values, in other words,
transition components between the ``plus'' and 
the ``minus''
edge states.

For the matrix representation of the Berry connection, it is convenient
to align the edge states to form a $2 \times 2$ matrix,
\begin{align}
M \equiv (\eta^+,\eta^-) \, .
\end{align}
Then the $2\times2$ Berry connection matrix (\ref{Ai}) is given by
\begin{align}
A_i = i \int_{-L}^L dx^4 M^\dagger \frac{d}{dp_i} M \, .
\end{align}
If we change the basis of the edge states in such a way that the ortho-normalization
condition (\ref{ortho}) is preserved, 
\begin{align}
\eta^a \rightarrow \eta^b V(p_i)_b^{\; a}
\end{align}
then the matrix $V$ needs to be 
a unitary matrix, $V \in$ U(2).
In terms of $M$, the gauge transformation acts as $M \to M V$.
The non-Abelian Berry connection (\ref{Ai}) is transformed as a gauge field, while the scalar operator $\Phi$
transforms as an adjoint representation scalar field,
\begin{align}
A_i \rightarrow i V^\dagger \frac{d}{dp_i} V + V^\dagger A_i V, 
\quad \Phi \rightarrow V^\dagger \Phi V.
\end{align}
Since this gauge transformation is merely a change of the basis of the edge states,
only the eigenvalues of the scalar field are gauge-invariant quantity
independent of the edge state basis.

Using the explicit edge states (\ref{edgenon2}) and (\ref{edgenon}), we can calculate
the Berry connection $A_i$ and the scalar $\Phi$. It turns out that choosing $V = U^\dagger$
provides a symmetric expression for the result. Using that basis~\footnote{
If we use the original basis, we obtain a diagonal expression for the scalar field,
$
\Phi = \left(L\coth(2 p L)-1/2p\right) \sigma_3$.
}, we obtain
\begin{align}
& A_i = \left(\frac{2 p L}{\sinh(2 p L)}-1\right)\frac{\epsilon_{ijk} p_k}{2p^2} \sigma_j \, , 
\\
& \Phi = \left(\frac{2 p L}{\tanh(2 p L)}-1\right) \frac{p_i}{2p^2}\sigma_i \, .
\end{align}
This Berry connection together with the matrix field $\Phi$
is identical to the well-known 't Hooft--Polyakov monopole~\cite{'tHooft:1974qc,Polyakov:1974ek} in the BPS limit \cite{Bogomolny:1975de,Prasad:1975kr}.

We find
that the Berry connection has a non-Abelian monopole charge,
\begin{align} 
1 = \frac{1}{4\pi} \int d^3p \; \frac12 \epsilon_{ijk} \; {\rm tr}\left[D_i \Phi F_{jk}\right]
\label{mag1}
\end{align}
Here we have defined
the covariant derivative and the field strength as usual,
\begin{align} 
&D_i \Phi \equiv \frac{\partial}{\partial p_i} \Phi - i [A_i, \Phi] , 
\\
&F_{ij} \equiv 
\frac{\partial}{\partial p_i} A_j 
-\frac{\partial}{\partial p_j} A_i
- i [A_i, A_j] \, . 
\end{align}
The monopole satisfies the famous BPS equation
\begin{align}
D_i \Phi = \frac12 \epsilon_{ijk} F_{jk} \, .
\end{align}
The 't Hooft--Polyakov monopole is a monopole solution of $SU(2)$ Yang--Mills theory
coupled to a scalar field $\Phi$ in the adjoint representation. We here have provided
an explicit example of the edge states whose topological property can be characterized
by the non-Abelian monopole.

The reason why we obtained the 't Hooft--Polyakov monopole is hidden in a parallelism to
the Nahm construction of monopoles. For a brief review of the Nahm construction, 
see Appendix~\ref{sec:Nahm_const}. The Nahm construction uses a Dirac zero mode of a certain Hamiltonian,
and our edge states satisfy exactly the same equation with exactly the same 
normalizability condition, (\ref{pipeta}) and (\ref{ortho}). So, as a result, it is required
that the Berry connection becomes that of the 't Hooft--Polyakov monopole.
Because the Nahm construction applies to not just the single monopole in $SU(2)$ gauge theory
but to broad species of non-Abelian gauge theories with arbitrary number of monopoles, 
we expect that this will lead to a mine of topological charges provided by edge states
in general.

If we make a trivial dimensional reduction by replacing $p_3$ by a mass $m_3$
as before, then we can think of the edge states as states in two dimensions.
The TKNN number for this set of edge states is provided by ${\rm tr}F_{12}$
(see \cite{hatsugai2006topological}).
However, since the non-Abelian monopole is that of $SU(2)$ gauge theory,
we find that the non-Abelian Berry connection has a trivial first Chern class:
${\rm tr} F_{12}=0$.

\begin{figure}[t]
\begin{center}
\includegraphics[width=8cm]{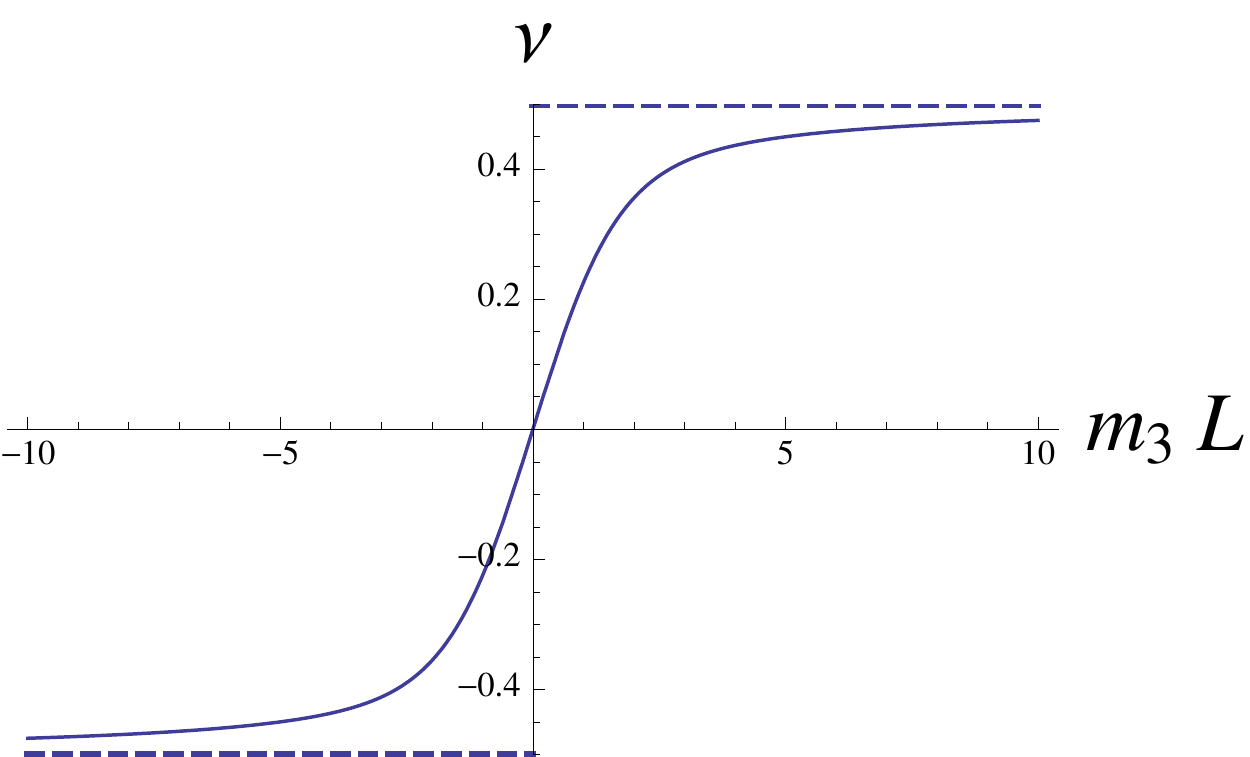}
\end{center}
\caption{The TKNN number (\ref{TKNN}) (dashed line) and our non-Abelian analogue (\ref{ana}) (solid line). The TKNN number is a step function of $m_3$, while our non-Abelian analogue is not singular. 
The asymptotic behavior is shared. In the limit $L\rightarrow \infty$, two lines coincide.}
\label{figntknn} 
\end{figure}

Nevertheless, we have another field strength which is invariant under the
$SU(2)$ gauge symmetry,
${\rm tr}[\Phi F_{12}]$. In fact, this invariant is nothing but the one providing the
non-Abelian monopole charge.
An explicit calculation gives
\begin{align}
&
\frac12 \epsilon_{ijk} {\rm tr}[\Phi F_{jk}] 
\nonumber 
\\
&
= \frac{-p_i}{p^4}
\left(1\!-\!pL \coth pL\right)\left(1\!-\!\left(\frac{pL}{\sinh pL}\right)^2\right)\, ,
\end{align}
which is integrated to provide (\ref{mag1})~\footnote{Note the relation 
$\epsilon_{ijk}\partial_i {\rm tr}[\Phi F_{jk}]=\epsilon_{ijk}{\rm tr}[D_i \Phi F_{jk}]$.}.
It would be instructive to calculate an analogue of the TKNN number (\ref{TKNN}).
Using this non-Abelian flux, one can compute an integral
\begin{align}
\tilde{\nu}
\equiv \frac{1}{4\pi}
\int dp_1 dp_2 \, {\rm tr}[\Phi F_{12}] \, .
\label{ana}
\end{align}
Since the 't Hooft--Polyakov monopole has a unit magnetic charge, it is easy to
observe 
\begin{align}
\lim_{m_3 \rightarrow \pm\infty} \tilde{\nu} = \pm \frac12 \, .
\end{align}
The difference from the TKNN number (\ref{TKNN}) is that the non-Abelian monopole
is not singular, and has a nonzero size $\sim 1/L$. In fact, the functional form of 
$\tilde{\nu}(m_3)$ is not a step function (which is the case for (\ref{TKNN})) but a
smooth function which interpolates $\pm 1/2$. For the explicit form, see Fig.~\ref{figntknn}.
In the limit $L \to \infty$, the 't Hooft--Polyakov monopole is reduced to the Dirac monopole, which is singular, and thus $\tilde{\nu} \to \nu$.

\subsection{Relation to domain-wall fermion in lattice QCD}

Our starting point, the 4D class A theory (\ref{4dtopH}), lives in a space with 
an extra dimension $x^4$. A similar technique is quite popular in lattice QCD where
chiral fermions in $1+3$ dimensions are introduced via a domain wall in the extra dimension,
called the domain-wall fermion formalism \cite{kaplan1992method,shamir1993chiral}.
See also~\cite{Kimura:2015ixh}.
Here let us discuss a difference between the domain-wall fermion and our class A topological insulator with boundaries.

Before getting to the lattice fermions, we here consider what is a possible boundary condition.
In the previous sections, we adopted a choice $M=\gamma_5$ (\ref{bcmin}).
However, more generally, the equation $\{M,\gamma_4\}=0$ 
(\ref{Mgamma4}) may have other solutions.
For example,
\begin{align}
M=\gamma_3 \, 
\label{bcmax}
\end{align}
which breaks the $SO(3)$ rotation symmetry.
Let us show that the choice is similar to a standard boundary condition in a 3D
Weyl semimetal \cite{Witten:2015aoa}. 
The Hamiltonian of the Weyl semimetal near the cone 
is given by ${\cal H}=\sigma_1 p_1 + \sigma_2 p_2 + \sigma_3 p_3$. 
Let us introduce a boundary
$x^1=0$, then the self-conjugacy condition $\{M,\sigma_1\}=0$ results in 
a solution such as $M=\sigma_3$ which breaks 
the momentum rotation symmetry $SO(2)$ in the $(p_2,p_3)$ space.
This resembles our choice (\ref{bcmax}) above. The important difference between this
3D Weyl semimetal and our 4D class A topological insulator
is that we have an alternative choice $M=\gamma_5$ (\ref{bcmin}) which does not spoil the
momentum rotation symmetry.

Now we are ready for discussing the difference from the lattice domain-wall fermions.
For the domain-wall fermions, one starts with a Dirac Hamiltonian
\begin{align}
H = \sum_{i=1,2,3} \gamma^0 \gamma_i p_i + \gamma^0\gamma_5 p_5 -  im\gamma^0\, 
\label{Hdf}
\end{align}
which can be derived from a Dirac equation in $1+4$ dimensions.
Remark that the Hamiltonian \eqref{4dtopH} is the same as the Hermitian operator used in the domain-wall/overlap formalism, which plays a role of the translation generator in the extra dimension.
But we are now dealing with the domain-wall fermion in the Hamiltonian formalism.
For the domain-wall fermion one introduces a wall at $x^5=0$, say. 
The domain-wall fermion is made by a change of the sign of the mass when one crosses the
wall. For example, one chooses $m>0$ for $x^5>0$ while $m<0$ for $x^5<0$. This mass profile
is understood as a chiral rotation, since if one applies the chiral rotation for the fermion $\psi \rightarrow \gamma_5 \psi$ then the Hamiltonian (\ref{Hdf}) changes as
\begin{align}
H \to \gamma_5 H \gamma_5 = 
\sum_{i=1,2,3}\gamma^0 \gamma_i p_i - \gamma^0\gamma_5 p_5 + im\gamma^0\, ,
\end{align}
which is equivalent to the $\gamma_5$-Hermiticity of the corresponding Dirac operator.
Note that the sign of the mass term changes, as well as the sign of the $p_5$ term.
So, the chiral rotation means the change of the sign of the mass at the same time as
the parity $x^5 \to -x^5$, which is equivalent to having the change of the sign of the mass
when one crosses the wall. The consistency of the fermion near the wall means $\gamma_5 \psi = \psi$, which is the chiral fermion. The massless chiral mode localizes at the wall.

Let us understand this domain-wall fermion in terms of our generic argument of
the boundary condition $\tilde{M}\psi = -\psi$ at $x^5=0$. 
From the Hamiltonian (\ref{Hdf}), a consistent boundary
condition needs $\tilde{M}$ satisfying $\{\tilde{M},\gamma^0\gamma_5\}=0$. The domain-wall fermion
formalism uses the choice $\tilde{M} = -\gamma_5$ for the Hamiltonian (\ref{Hdf}), because the lattice QCD does not like to break the $1+3$-dimensional Lorentz invariance. 

We can find a relation to our topological insulator.
Noting that $i\gamma^0$ and $\gamma^0\gamma_i$ in (\ref{Hdf}) 
are Hermitian ($\gamma^0$ is anti-Hermitian itself) and satisfy the Euclidean Clifford algebra, we can actually relabel the indices of (\ref{Hdf}) and see the equivalence to
our 4D topological insulator (\ref{4dtopH}):
\begin{align}
\gamma^0\gamma_i \rightarrow \gamma_i, \quad
\gamma^0 \gamma_5 \rightarrow \gamma_4,
\quad
-i \gamma^0 \rightarrow \gamma_5 \, .
\label{corrgamma}
\end{align}
In our terminology, using (\ref{corrgamma}), the domain-wall fermion corresponds to $M = i \gamma_5 \gamma_4$, while our boundary condition is $M=\gamma_5$.
In other words, in the terminology of the lattice domain-wall fermion, the domain wall boundary condition is $\tilde{M}=-\gamma_5$, while our boundary condition is $\tilde{M}=-i\gamma^0$.
This signals an important difference between our boundary condition and the domain-wall fermion:
The domain-wall fermion $\tilde{M}=-\gamma_5$ has the same index as the wall position $x^5=0$,
while our choice $\tilde{M}=-i\gamma^0$ does not. Generically, in the domain-wall formulation
of lattice QCD, when one has two walls, they produce a pair of $\tilde{M}=-\gamma_5$ and
$\tilde{M}=\gamma_5$ to end up with vector-like fermions. However in our topological insulator,
this pairing does not apply, since our boundary condition $\tilde{M}=-i\gamma^0$ is not
related to the coordinate $x^5$. That is why we can choose in fact the same boundary conditions
at the two boundaries.

For our topological insulators, we can choose freely boundary conditions at each boundary respectively~\footnote{This is at least for the class A system, which does not possesses any discrete symmetries. The situation will be changed if we consider other classes with additional symmetries.}.
A generic choice of the boundary conditions will reveal how universal our topological charges of the edge states are.
It would be an interesting future work.

%

%
%
%
%


\section{Magnetic field, non-commutative space and D-brane}
\label{sec5}

In this section we consider how the non-Abelian monopole is deformed once
we apply a magnetic field on the 4D class A system.
It turns out that the non-Abelian monopole still persists, and it is identified
as a monopole in a non-commutative space~\cite{Hashimoto:1999zw,Gross:2000ss,Gross:2000ph,Gross:2000wc}~\footnote{See \cite{Douglas:2001ba} for a review of field theories on non-commutative spacetime}. 
Since the properties 
of such a monopole in non-commutative space 
can be analyzed by 
a D-brane construction in string theory, we use the D-brane interpretation
to explore the properties of edge states. We find that the tilted D-brane configuration
clarifies the shift of the fermion momentum for the edge states.

We start with the 4D class A system (\ref{4dtopH}), and consider the following
``magnetic'' field in four dimensions,
\begin{align}
F_{12}= F_{34}=B
\end{align}
while the other components are set to zero. $B(>0)$ is a constant field strength which is self-dual
in the 4D space. This choice of the field strength is a typical configuration for the 4D quantum Hall effect~\cite{Zhang:2001xs,Qi:2008ew}, and the simplest for having a consistent BPS equation satisfied by the monopole, as we will see.

Due to the magnetic field, the momenta are now non-commutative
to each other,
\begin{align}
[\hat{p}_1,\hat{p}_2] = [\hat{p}_3,\hat{p}_4]=i B\,.
\label{sdB}
\end{align}
The 't Hooft--Polyakov monopole defined by the edge states 
lives in the space spanned by $(p_1,p_2,p_3)$, thus 
we are now looking for a monopole in a non-commutative momentum space $[p_1,p_2]=iB$.
Any function in 
the non-commutative space can be expanded by a creation-annihilation operator
\begin{align}
\hat{a} \equiv \frac{1}{\sqrt{2B}}(\hat{p}_1 + i \hat{p}_2)\, , \quad
\hat{a}^\dagger  \equiv \frac{1}{\sqrt{2B}}(\hat{p}_1 - i \hat{p}_2)\, 
\end{align}
satisfying $[\hat{a},\hat{a}^\dagger]=1$.
The edge states need to satisfy, from (\ref{eq1}) with $\xi=0$ at $\epsilon = -m$,
\begin{align}
\left(
\bar{e}_1 \hat{p}_1 +\bar{e}_2 \hat{p}_2
+\bar{e}_1 (p_3 + B x^4)
 - i \frac{d}{dx^4}
\right)
\eta = 0 \, .
\end{align}
Note here that we have defined $p_3\equiv \partial_3$ (without the hat) such that the explicit
magnetic-field dependence in the $\hat{p}_3$-$\hat{p}_4$ space 
can be seen as the $+Bx^4$ term.

According to the Nahm construction of monopoles in non-commutative space
\cite{Gross:2000ss,Gross:2000ph,Gross:2000wc}, this equation
is exactly the one to solve for the construction of all solutions satisfying the BPS
monopole equation in the non-commutative space,
\begin{align}
D_i \Phi = \frac12 \epsilon_{ijk} F_{jk} \, ,
\end{align}
where the scalar field $\Phi$ and the gauge field $A_i$ are functions of the non-commutative
coordinates $(\hat{p}_1,\hat{p}_2,p_3)$. Therefore, we conclude that 
putting a self-dual magnetic field (\ref{sdB}) in the class A topological insulator in four dimensions
with one/two boundary leads to a BPS Abelian/non-Abelian monopole in a non-commutative 
(momentum) space.

The D-brane interpretation of the monopole \cite{Hashimoto:1999zw} is provided
by a slanted D1-brane stuck to D3-brane(s). In particular, when we have a single
boundary surface, the monopole is that of a $U(1)$ gauge theory, which is, a
non-commutative Dirac monopole. The explicit solution was given in \cite{Gross:2000ss}
which exhibits an interesting behavior   
\begin{align}
\langle 0 | \Phi | 0 \rangle \sim \frac{1}{B} p_3 
\label{0p0}
\end{align}
for $p_3 \rightarrow +\infty$. Here $|n\rangle$ ($n=0,1,2,\cdots$) is the Landau level,
that is, a basis of the Fock
space spanned by the operator $\hat{a}^\dagger$ where $\hat{a}|0\rangle=0$, and any
function in the non-commutative space can be spanned by $|n\rangle\langle m |$.
In Eq.~(\ref{0p0}) we look at the lowest Landau level for simplicity. 
Eq.~(\ref{0p0}) means that the location $\Phi$ of the fermion for given $p_3$ at the lowest Landau
level is linearly dependent for large positive $p_3$. For larger $p_3$, the fermion wave function
on the edge state goes deeply inside the bulk away from the boundary, linearly.
Since the scalar field $\Phi$ is nothing but the D-brane shape in string theory, 
the configuration (\ref{0p0})
was interpreted as a slanted D1-brane~\footnote{The exact shape can be understood only 
in a gauge-invariant quantity 
\cite{Hashimoto:1999az,Mateos:2000qq,Hashimoto:2000mt,Hashimoto:2000kq,Hashimoto:2001pc}
such as a scalar configuration after the so-called Seiberg--Witten
map \cite{Seiberg:1999vs}.}.
So the shape of the D1-D3-brane system provides the information of the location of
the edge state fermions.


\section{Conclusion and discussions}
\label{sec6}

In this paper, we find that the edge states of a four-dimensional topological insulator of class A
have topological charges. For a single boundary with a certain boundary condition, 
a Dirac monopole in momentum space emerges. Upon a dimensional reduction, it is identical
to the TKNN number. When there are two parallel boundaries, the topological charge is
identified as that of a BPS 't Hooft--Polyakov monopole in an $SU(2)$ gauge theory.
It defines a non-Abelian analogue of the TKNN number.

The classification of topological insulators is deeply concerned with dimensions,
and the dimensional reduction technique \cite{Qi:2008ew,Ryu:2010zza} is widely used for analyses. Here
we propose another way to change the dimensionality, via introducing boundaries and considering
topological nature of the edge states.
We remark that this topological property of the edge state might be related to the surface topological order, which appears in the interacting topological insulators~\cite{Wang:2013uky,Bonderson:2013:JSM,Wang:2013zja,Chen:2013jha,Metlitski:2015PRB}.
See also a recent article~\cite{Seiberg:2016rsg}.
It would be interesting to study more explicit connection to such an argument.

In the study of topological systems, the open boundary condition has been typically applied to observe the edge state so far.
Since varieties of the boundary conditions are now allowed, it would be interesting to explore 
all possible boundary conditions.
As an example, we have explained that the domain-wall fermions in lattice QCD
corresponds to a different boundary condition. Furthermore,
once more than two boundaries are introduced, more exotic non-Abelian 
examples are expected to appear, such as $SU(n)$ monopoles. In that case,
the added interior ``boundaries'' 
can be interpreted as a surface junction of multilayer systems.
Consistent boundary conditions may be classified by K-theory, as in the case of the bulk topological properties because finding the boundary condition matrix, e.g. \eqref{Mgamma4}, seems a matter of the Clifford algebra.
Exhausting all possible boundary conditions associated with the edge topological numbers is an important future direction.

In the last section we demonstrated that the technology of D-brane engineering in string theory
is useful for extracting information of fermions. The location of the edge-state fermions in
the momentum space can be identified with the shape of the D-brane. We have made in our previous paper \cite{Hashimoto:2015dla} that the shape of the fermion dispersion of topological insulators 
can be interpreted as the shape of a particular D-brane, so it would be interesting to
further explore the relation between topological charges of edge states and D-branes in
string theory.

In the end, let us discuss how to realize our proposal in experiments.
Our model is the class A system, showing the 4D quantum Hall effect, which could be realized using ultracold atoms~\cite{Price:2015qca}.
Furthurmore, as mentioned before, our analysis is also applicable to the 3D chiral topological insulator (class AIII) \cite{Hosur:2010PRB}, which is connected with the 4D class A through the standard dimensional reduction.
Thus, imposing the boundary condition studied in this paper for these systems, we could observe the flat band dispersion at the surface, as a signal of the topological nature of the edge state.
In order to discuss such a realization in experiments, it will be required to construct some lattice model exhibiting the fermion boundary conditions which we adopted.
In particular, a lattice model having
the non-Abelian TKNN number would be of importance. 

%
%
%

%


\subsection*{Acknowledgments}
K.~H.~would like to thank S.~Yamaguchi, H.~Fukaya and T.~Onogi for valuable discussions.
The work of K.~H.~was supported in part by JSPS KAKENHI Grant Number 15H03658 and 15K13483.
The work of T.~K.~was supported in part by JSPS KAKENHI Grant Number 13J04302.


\vspace{1em}

\appendix

\section{Review of Nahm construction of BPS monopoles}
\label{sec:Nahm_const}

The Nahm construction \cite{Nahm:1979yw} 
of monopoles, or the Nahm construction in short, is a way to obtain all solutions of 
the BPS monopole equation
\begin{align}
D_i \Phi  = \frac12 \epsilon_{ijk} F_{jk}
\label{bpsn}
\end{align}
for a Yang--Mills--Higgs theory with a non-Abelian gauge group in three spatial dimensions.
(For a review, see \cite{corrigan1979self,corrigan1980some,prasad1980instantons}.)
Here we briefly review the Nahm construction and its generalization to the one in a non-commutative space~\footnote{The D-brane interpretation of 
the Nahm equation was given in \cite{Diaconescu:1996rk}.
For the D-brane interpretation of the Nahm construction itself, see
\cite{Hashimoto:2005yy}.}.

The Nahm construction consists of the following three steps:
\begin{itemize}
\item[1)]
For $k$-monopoles, 
solve the Nahm equation for $k\times k$ Hermitian matrices $T_i (i=1,2,3)$ as a function
of a parameter $\xi$:
\begin{align}
\frac{d}{d\xi} T_i = i\epsilon_{ijk} T_j T_k\, .
\label{Nahm}
\end{align}
The $\xi$ space is defined on a period $-s<\xi<s$ for $SU(2)$ monopoles.
Note that for a single monopole $k=1$ the Nahm equation can be solved trivially
by $T_i(\xi)=0$. 

\item[2)]
Solve a zero-mode equation
\begin{align}
\nabla^\dagger v(\xi) = 0
\label{DiracNahm}
\end{align}
for $v(\xi;x^i)$ where $\nabla^\dagger$ is defined as
\begin{align}
\nabla^\dagger \equiv i \frac{d}{d\xi} + i \sigma_i \left(x^i-T_i(\xi)\right) \, .
\end{align}
Here $\sigma_i (i=1,2,3)$ is the Pauli matrix, and the solutions $v^{(a)}(\xi)$ where
$a=1,2$ 
need to be normalized as
\begin{align}
\int_{-s}^s d\xi \; (v^{(a)})^\dagger v^{(b)} = \delta_{ab} . 
\end{align}

\item[3)]
Calculate the gauge field and the scalar field by the formulas
\begin{align}
&\Phi^{ab}(x) \equiv \int_{-s}^s \!d\xi \; v^{(a)\dagger} \xi v^{(b)} \, , 
\\
& A_i^{ab}(x) \equiv  \int_{-s}^s \! d\xi \; v^{(a)\dagger} i\frac{d}{dx^i} v^{(b)} \, .
\label{PhiAformula}
\end{align}
\end{itemize}
Then the defined gauge field and the scalar field satisfy the BPS monopole equation (\ref{bpsn}).

In the non-commutative space $[x^1,x^2]=i\theta$, the Nahm construction is only modified at the
Nahm equation \cite{Gross:2000ss}
\begin{align}
\frac{d}{d\xi} T_i +  \theta \delta_{i3}= i\epsilon_{ijk} T_j T_k \, .
\end{align}
Following the same procedures above with care on the operator orderings, one arrives at
BPS monopole solutions in the non-commutative space.

%

\bibliography{monopole-revtex}

\end{document}